\documentclass[preprint,12pt,authoryear]{elsarticle}

\usepackage[utf8]{inputenc}
\usepackage{graphicx}
\usepackage{amssymb}
\usepackage{amsmath}
\usepackage{booktabs}
\usepackage{float}
\usepackage{chngcntr}
\usepackage{hyperref}

\biboptions{round}

\journal{Journal of Biomechanics}

\begin{document}

\begin{frontmatter}

\title{Sleep-related forcing of cerebrospinal fluid streaming transport through the third ventricle}

\author[inst1]{Pengcheng Zhao}
\author[inst1]{Luke McTavish}
\author[inst1,cor1]{Christoph Bruecker}
\cortext[cor1]{Corresponding author.}
\ead{Christoph.Bruecker@citystgeorges.ac.uk}

\affiliation[inst1]{
organization={City St George's, University of London},
addressline={Northampton Square},
city={London},
postcode={EC1V 0HB},
country={United Kingdom}
}

\begin{abstract}
Quasi-periodic pressure pulses associated with cardiac and respiratory activity generate reciprocating cerebrospinal fluid (CSF) motion in the human ventricles. While weak and more regular during wakefulness, oscillatory fluid motions have recently been found to be much more exaggerated during NREM sleep, appearing as large flushes in successive pulse trains. The present study investigates whether this forcing can induce greater residual streaming through the human ventricles. Transient CFD simulations were performed in the third ventricle under awake and sleep-related flow pulses. Sleep-related forcing produced clear caudal and rostral preferential pathways and less overall recirculation in the cavity. The streaming flux relative to the stroke volume increased from a low value of about \(0.540\%\) to \(28.4\%\) during sleep. Thus, sleep-related forcing changes not only the magnitude of reciprocating fluid displacement but also the relative contribution of residual streaming. These findings demonstrate differences in hydrodynamic organisation in the third ventricle and highlight the importance of this organisation for intraventricular fluid exchange. 
\end{abstract}


\end{frontmatter}

\section{Introduction}

Cerebrospinal fluid (CSF) motion in the ventricular system contributes to mechanical protection and fluid exchange around the brain \citep{sakka2011anatomy,iliff2012paravascular}. In addition to the quasi-steady transport of CSF produced in the ventricles, CSF motion also has an oscillatory component, which is driven primarily by quasi-periodic pressure pulses associated with cardiac pulsation and respiration \citep{yamada2013influence,chen2015dynamics}. Typically, the net CSF flow rate due to production and drainage in the system is one order of magnitude lower than the volume that is displaced caudally and rostrally in each cycle \citep{spijkerman2019phase}. The resulting quasi-periodic flow reversals make ventricular CSF motion a candidate for so-called steady streaming \citep{haselton1982flow,khani2018anthropomorphic}, namely, the generation of residual mean motion that contributes to fluid exchange even when the net fluid displacement over a cycle is close to zero. Oscillatory, reciprocating flows can therefore still generate effective transport and mixing even when the cycle-averaged net flow is small. Such residual motion can arise when an oscillatory flow interacts with complex geometries and spatially non-uniform velocity gradients; therefore, its organisation cannot be inferred from net flow alone. As an example, streaming has been demonstrated as an important transport mechanism in reciprocating flow through bifurcations in the human airways \citep{haselton1982flow,wanigasekara2024mean}.

The focus of our study herein is the region of the third ventricle, as it occupies a central position in the CSF passage: it receives CSF from the lateral ventricles through two small channels, the FOM, and communicates with the fourth ventricle through another narrow channel, the cerebral aqueduct. Its curved walls, local anatomical constraints such as the position and size of a possible interthalamic adhesion (see \cite{duqueparra2022interthalamic}), and the shape of the FOM and aqueduct can organise the hydrodynamic pattern of the residual streaming motion. The constrictions of the FOM and the aqueduct can also amplify local velocity gradients relative to those within the main ventricular cavity. The third ventricle therefore provides an important setting in which to examine how physiological forcing is converted into cross-ventricular exchange and residual streaming.

Previous CFD studies have already been conducted to analyse CSF flow in the third ventricle \citep{kurtcuoglu2007mixing,kurtcuoglu2007computational}. These early studies used subject-specific models of the third ventricle and examined aqueductal jets, recirculation, and laminar flow characteristics; however, they used boundary conditions relevant only to the awake condition. In the meantime, the ground-breaking paper by Fultz et al. has highlighted the fundamentally different situation during NREM sleep \citep{fultz2019coupled}. In their human neuroimaging study, they showed that non-rapid eye movement (NREM) sleep is associated with large-volume, slow CSF oscillations coupled to neural and haemodynamic activity \citep{fultz2019coupled}. These events differ substantially in amplitude and time scale from the smaller, faster pulsations that characterise oscillatory CSF exchange during wakefulness. This led Fultz et al. to call this the "Washing Machine", which helps to clear the brain. Animal work has indeed shown that sleep enhances metabolite clearance from the brain \citep{xie2013sleep}. 

The current work incorporates these new observations into revised boundary conditions for the CFD simulations. It focuses on the hydrodynamic question of whether sleep-related CSF oscillations alter local ventricular flow organisation, exchange, and streaming relative to an awake condition. CFD provides a complementary approach for resolving local ventricular flow structures under controlled physiological conditions. A representative human third-ventricle geometry was used to compare awake and sleep-related oscillatory forcing while the geometry, mesh, and material properties were held fixed. The analysis considered midsagittal velocity fields, preferential pathways, stroke volume, the residual streaming flux \(Q_r\), and the streaming efficiency. The objective was to determine whether sleep-related forcing changes both the magnitude of CSF exchange and the spatial organisation of streaming transport through the third ventricle. This controlled design attributes differences in the computed response to the imposed forcing amplitude and period.


\section{Methods}

\subsection{Geometry and mesh generation}

A representative human third-ventricle geometry was obtained from the public 3D model platform Thingiverse (published by the user akshay\_d21, see https://www.thingiverse.com/thing:2199907). According to the model description, the geometry was derived from BodyParts3D/Anatomography. It was transferred into Fusion 360 to generate a computational fluid domain that contains the third-ventricle body, the bilateral FOM, and the cerebral aqueduct (aqueduct of Sylvius). The reconstructed fluid-domain volume was approximately \(1.25\,\mathrm{mL}\), which is within the range of adult third-ventricle volumes reported by \citet{hernandezcortes2023third}. The inferior end of the cerebral aqueduct was extended further into a circular pipe with small taper to ensure smooth entrance into the aqueduct when the flow is rostral. The equivalent diameter of the inferior end with a cross-sectional area of \(A_{\mathrm{Aq}}=2.86\,\mathrm{mm^2}\) results to \(D_{\mathrm{Aq}}=1.9\,\mathrm{mm}\), which lies within the range reported for the aqueduct diameter in normal subjects \citep{oner2017quantitative}. The combined area of the two FOM inlets was \(A_{\mathrm{in}}=11.24\,\mathrm{mm^2}\).

\begin{figure*}[t]
\centering
\includegraphics[width=0.85\textwidth]{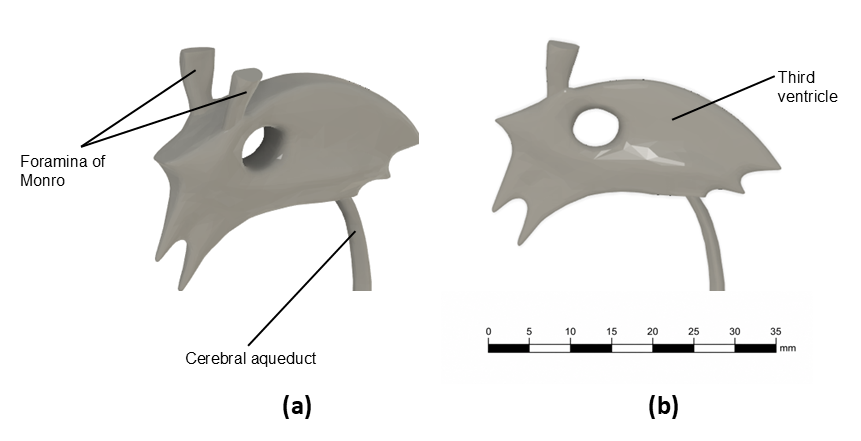}
\caption{Reconstructed third ventricle geometry used for CFD simulation, adapted from the public Thingiverse model by akshay\_d21. 
(a) Three-dimensional view of the processed fluid domain showing the FOM and the cerebral aqueduct. 
(b) Lateral view of the same geometry with scale bar and outlet extension. The vertical axis of the coordinate system is aligned with the axis of the inferior end of the aqueduct.}
\label{fig:geometry}
\end{figure*}

The FOM were defined as inlet boundaries, while the extended cerebral aqueduct was defined as the outlet boundary. All remaining surfaces were treated as no-slip walls.
 
The processed geometry was imported into ANSYS Meshing, where an unstructured mesh was generated automatically using a global element size of \(9\times10^{-5}\,\mathrm{m}\) together with local refinement around the FOM, the cerebral aqueduct, and strongly curved regions. Named selections were assigned to the inlet, outlet, and wall boundaries before import into ANSYS Fluent. The same geometry and baseline mesh, comprising approximately \(4.84\) million elements, were used for the awake and sleep simulations so that differences between cases reflected the prescribed forcing rather than geometry or discretisation. A mesh-independence study was carried out and showed sufficient resolution for both cases; see the documentation in ~\ref{app:independence}.

\subsection{Fluid model and numerical method}

Cerebrospinal fluid (CSF) was modelled as an incompressible Newtonian fluid with water-like properties with the density \(\rho=1000\,\mathrm{kg\,m^{-3}}\) and the kinematic viscosity \(\nu=1.0\times10^{-6}\,\mathrm{m^2\,s^{-1}}\). Numerical simulations were performed in ANSYS Fluent using a pressure-based solver for the transient incompressible Navier--Stokes equations. The Reynolds-number estimates given later in Section 3.2 remained below the conventional transition range for internal flow; a laminar formulation was therefore used, with rigid no-slip walls. Each case was simulated for three complete cycles, and the final cycle was used for flow-metric extraction after the initial transient response. Residuals of the continuity equation and velocity components were monitored, and the outlet flow-rate waveform was checked for temporal stability and periodicity.

\subsection{Boundary conditions and simulation settings}

The inlet velocity was prescribed as a spatially uniform sinusoidal waveform on both sides of the FOM,

\begin{equation}
U(t)=U_{\mathrm{peak}}\sin\left(\frac{2\pi t}{T}\right),
\label{eq:sinusoidal_velocity}
\end{equation}

where \(U_{\mathrm{peak}}\) is the peak inlet velocity and \(T\) is the oscillation period. The resulting Q-t flow waveform has a positive part that generates caudal flow from the third ventricle towards the fourth ventricle, whereas the negative part denotes reverse flow in the rostral direction.

Herein, we take into account only the oscillatory part of CSF transport for both the awake and sleep conditions in order to isolate the streaming effect induced by the flow reversals. This is justified because the net CSF flow rate due to production and drainage in the system is one order of magnitude lower than the volume that is displaced caudally and rostrally in each cycle \citep{spijkerman2019phase}. As a first-order approximation, we further represent the successive flow pulses in both conditions using an idealised sinusoidal waveform, with each pulse flushing the same volume of fluid (the stroke volume \(SV\)) caudally and rostrally. This assumption is at least reasonable for the awake condition; see \citep{gholampour2021boundary}. So far, no detailed flow measurements in the aqueduct have been possible during sleep. In the absence of such measurements, the first-order approximation is a reasonable simplification for focusing on the streaming effect under similar boundary conditions; see also \citep{beggs2019predicting}. This allowed the comparison to isolate the effects of amplitude and oscillation period on the streaming effect. 

In the following, the reference velocity inside the aqueduct is introduced using the relevant cross-sectional areas. For a sinusoidal inlet waveform, the positive-half-cycle stroke volume in the model was related to the peak inlet velocity by

\begin{equation}
SV=\int_{0}^{T/2}Q(t)\,\mathrm{d}t
  =\frac{A_{\mathrm{in}}U_{\mathrm{peak,in}}T}{\pi}
  =\frac{A_{\mathrm{Aq}}U_{\mathrm{peak,Aq}}T}{\pi}.
\end{equation}

For the awake condition, the target stroke volume was taken from the data reported by \citet{markenrothbloch2018}, giving \(SV_{\mathrm{awake}}=6.0\,\mu\mathrm{L}\). With \(T_{\mathrm{awake}}=1.0\,\mathrm{s}\), this gave the FOM inlet amplitude \(U_{\mathrm{peak}}=0.0017\,\mathrm{m\,s^{-1}}\). The awake simulation used a solver time step of \(0.02\,\mathrm{s}\), a total physical time of \(3.0\,\mathrm{s}\), and three complete cycles.

For the sleep condition, the forcing amplitude and period were approximated from the large, spatially coherent CSF inflow-related fMRI signal changes reported during NREM sleep by \citet{fultz2019coupled}. The reported data showed large waves of CSF inflow approximately every \(20\,\mathrm{s}\). A detailed inspection of their signals revealed that each wave consisted of 3--4 successive larger oscillatory peaks with a peak-to-peak interval of \(T \approx 4\,\mathrm{s}\); these observations suggested a representative period of \(T_{\mathrm{sleep}}=4.0\,\mathrm{s}\), as probably driven by the breathing cycle during sleep. The estimated velocity amplitude, \(U_{\mathrm{peak,in}}=0.17\,\mathrm{m\,s^{-1}}\), is detailed further in ~\ref{app:sleep_velocity_estimation}. The corresponding nominal model stroke volume was approximately \(2.43\,\mathrm{mL}\). Relative to the reconstructed third-ventricle volume of approximately \(1.25\,\mathrm{mL}\), these values corresponded to approximately \(0.49\%\) of the ventricular volume per awake cycle and approximately \(1.9\) ventricular volumes per sleep cycle. The sleep simulation used a solver time step of \(5.0\times10^{-4}\,\mathrm{s}\), a total physical time of \(12.0\,\mathrm{s}\), and three complete cycles. Export was performed every 80 solver time steps, giving an export interval of \(0.04\,\mathrm{s}\) and 100 exported points per cycle. 

\begin{table}[H]
\centering
\begin{tabular}{lcccc}
\toprule
Case & \(U_{\mathrm{peak}}\) & \(T\) & Solver time step & Cycles \\
\midrule
Awake 
& \(0.0017\,\mathrm{m\,s^{-1}}\) 
& \(1.0\,\mathrm{s}\) 
& \(0.02\,\mathrm{s}\) 
& 3 \\

Sleep 
& \(0.17\,\mathrm{m\,s^{-1}}\) 
& \(4.0\,\mathrm{s}\) 
& \(5.0 \times 10^{-4}\,\mathrm{s}\) 
& 3 \\
\bottomrule
\end{tabular}
\caption{Inlet velocity boundary conditions and temporal settings for the awake and sleep simulations.}
\label{tab:inlet_velocity_settings}
\end{table}

A representative Courant--Friedrichs--Lewy (CFL) number was also estimated using \(C=U_{\mathrm{peak}}\Delta t/\Delta x\), with \(\Delta x=9\times10^{-5}\,\mathrm{m}\) taken as the representative mesh length scale \citep{courant1928partiellen}. The representative CFL numbers were \(C_{\mathrm{awake}}\approx0.38\) and \(C_{\mathrm{sleep}}\approx0.94\), indicating adequate temporal resolution for the selected time-step sizes.

\subsection{Characteristic flow parameters}

The Reynolds number based on the cerebral aqueduct diameter as the characteristic length scale is defined as

\begin{equation}
Re=\frac{U_{\mathrm{Aq,peak}}D_{\mathrm{Aq}}}{\nu},
\label{eq:reynolds_number}
\end{equation}

where \(U_{\mathrm{Aq,peak}}\) is the peak cross-sectionally averaged velocity in the cerebral aqueduct and \(D_{\mathrm{Aq}}\) is its equivalent diameter. The estimated Reynolds numbers were \(Re_{\mathrm{awake}}=12.7\) and \(Re_{\mathrm{sleep}}=1270\). Both values remained below the conventional transition range for internal pipe flow.

The Womersley number was used to characterise the relative importance of unsteady inertial effects and viscous diffusion in the oscillatory CSF flow \citep{womersley1955method}:

\begin{equation}
\alpha =
\frac{D_{\mathrm{Aq}}}{2}
\sqrt{\frac{2\pi}{T\nu}} .
\label{eq:womersley_number}
\end{equation}

The resulting values were \(\alpha_{\mathrm{awake}}\approx2.38\) and \(\alpha_{\mathrm{sleep}}\approx1.19\). Please note that these values refer to the flow in the aqueduct and are not representative of the typical scales in the ventricle.

\subsection{Post-processing and flow metrics}

Post-processing was performed in MATLAB using the exported transient velocity fields from ANSYS Fluent. Quantitative flow metrics were calculated from the cerebral aqueduct outlet; the midsagittal plane was used only for velocity-contour and streamline visualisation. Unless otherwise stated, flow metrics were evaluated over the final simulated cycle for each case.  

The net-streaming flow pattern becomes visible by time-averaging the velocity field which is \(\widehat{\mathbf{U}}_s\), called the mean streaming velocity field. This flow field is later analysed in detail using streamline patterns and contour plots. From this, the component of residual streaming flux across the boundary of the inferior end of the aqueduct is calculated as follows (see also \citet{wanigasekara2024mean}):  

\begin{equation}
Q_r
=
\int_{A}
\left[
\widehat{\mathbf{U}}_s\cdot
\mathbf{n}_{\mathrm{Aq}}
\right]
H\left(
\widehat{\mathbf{U}}_s\cdot
\mathbf{n}_{\mathrm{Aq}}
\right)
\,\mathrm{d}A,
\label{eq:qr_definition}
\end{equation}

where \(H\) is the Heaviside step function. The vector \(\mathbf{n}\) is the normal vector at the surface of the aqueductal cross-section. The streaming efficiency was then defined as

\begin{equation}
SE=\frac{Q_r \cdot T}{SV}.
\label{eq:stream_effect}
\end{equation}

Because of the annular structure of the aqueduct, the results of the streaming flux at the inferior end are mostly representative also for the upper cross-sections in the aqueduct. The aqueduct was selected for defining SE as it indicates how streaming is also affecting the fluid exchange at the boundary to the 4th ventricle, thus showing the importance of cross-ventricular exchange between the interconnected cavities of the brain.   

\section{Results}

\subsection{Instantaneous bidirectional flow over the cycle}

\begin{figure}[H]
\centering
\includegraphics[width=\textwidth]{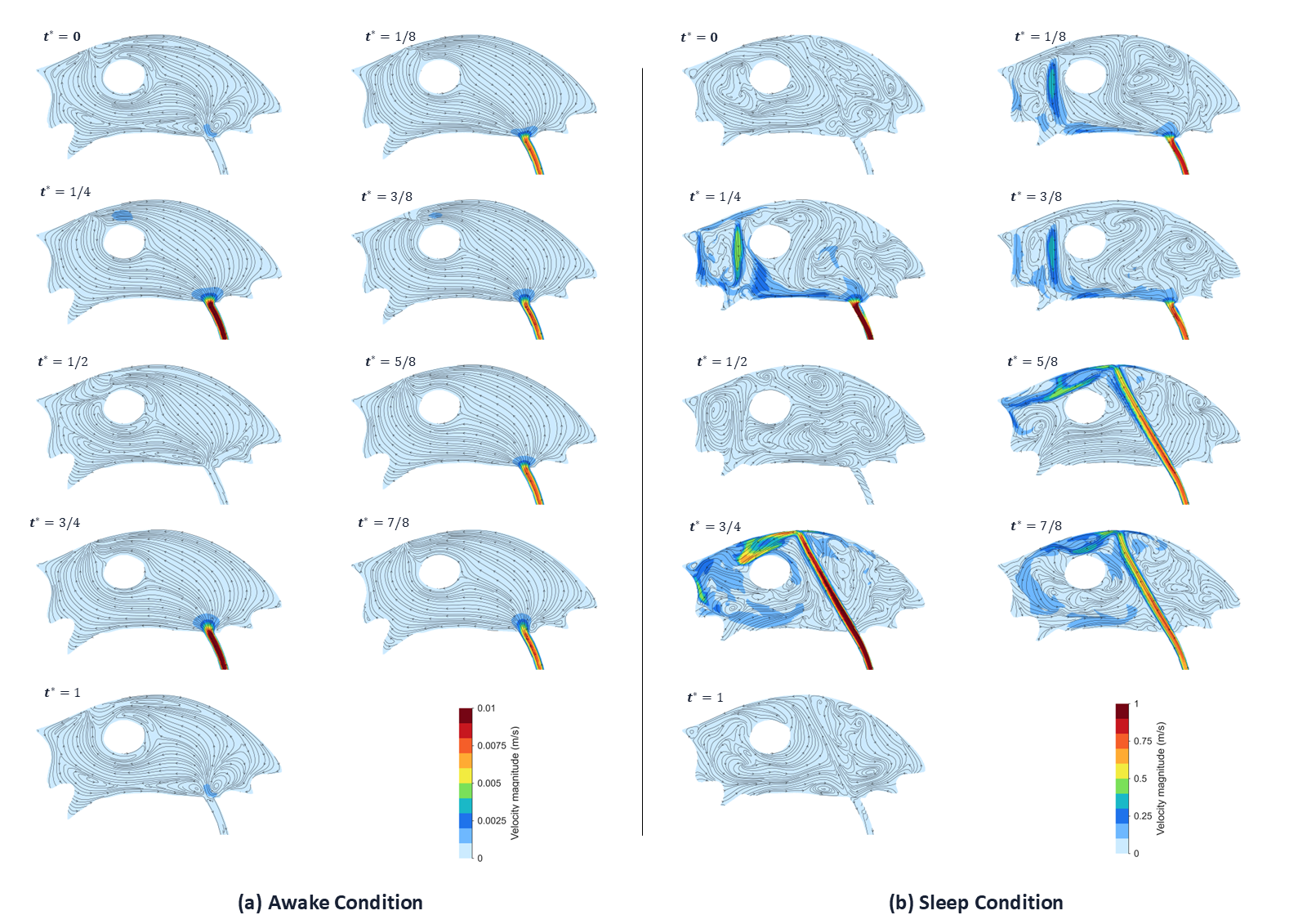}
\caption{Instantaneous midsagittal velocity magnitude and streamline distributions at nine equally spaced phases with \(\Delta t^* = 1/8\) during the final simulated cycle for (a) the awake condition and (b) the sleep condition. Here, \(t^*=(t-t_0)/T\). Separate colour-bar ranges are used because of the large difference in velocity magnitude between the two conditions.}
\label{fig:instantaneous_frames}
\end{figure}

The instantaneous velocity fields were compared at nine equally spaced phases of the final simulated cycle (Figure~\ref{fig:instantaneous_frames}). The normalised phase was defined as \(t^*=(t-t_0)/T\), where \(t_0\) is the start of the analysed cycle and \(T\) is the oscillation period. Thus, \(t^*=0\) and \(t^*=1\) represent equivalent phases in successive cycles. Under the prescribed sinusoidal convention, \(0<t^*<1/2\) corresponds to caudal flow from the FOM towards the cerebral aqueduct, whereas \(1/2<t^*<1\) corresponds to rostral/cephalad reverse flow. The phases \(t^*=0\), \(1/2\), and \(1\) are associated with flow reversal, at which the imposed inlet velocity approaches zero.

In the awake condition, the instantaneous motion remained relatively weak and diffuse throughout the cycle. The largest velocity magnitudes occurred near the cerebral aqueduct, particularly around the peak caudal and rostral phases at \(t^*=1/4\) and \(t^*=3/4\), respectively. Overall, the streamline patterns at peak inflow and outflow looked very similar, differing only in flow direction, meaning that the broad streamline topology recurred with the flow direction reversed. Most of the main third-ventricle cavity remained at a low velocity magnitude, and the instantaneous streamline patterns changed direction without forming a strongly concentrated pathway through the entire cavity.

In the sleep condition, both half-cycles produced substantially stronger and more spatially distinct motion patterns. During the caudal half-cycle, a concentrated pathway extended from the superior ventricular region towards the cerebral aqueduct, curving around the adhesion on the left side. During the rostral half-cycle, the flow reversed and a pronounced jet-like structure extended from the aqueductal region into the main ventricular cavity, passing the adhesion on the right-hand side. Stronger streamline curvature and more evident local recirculation accompanied this jet. Across the cycle, the broad streamline topology under sleep-related forcing showed greater spatial variation in regions of high velocity magnitude and clearer spatial organisation.

\subsection{Cycle-averaged streaming organisation}

\begin{figure}[H]
\centering
\includegraphics[width=\textwidth]{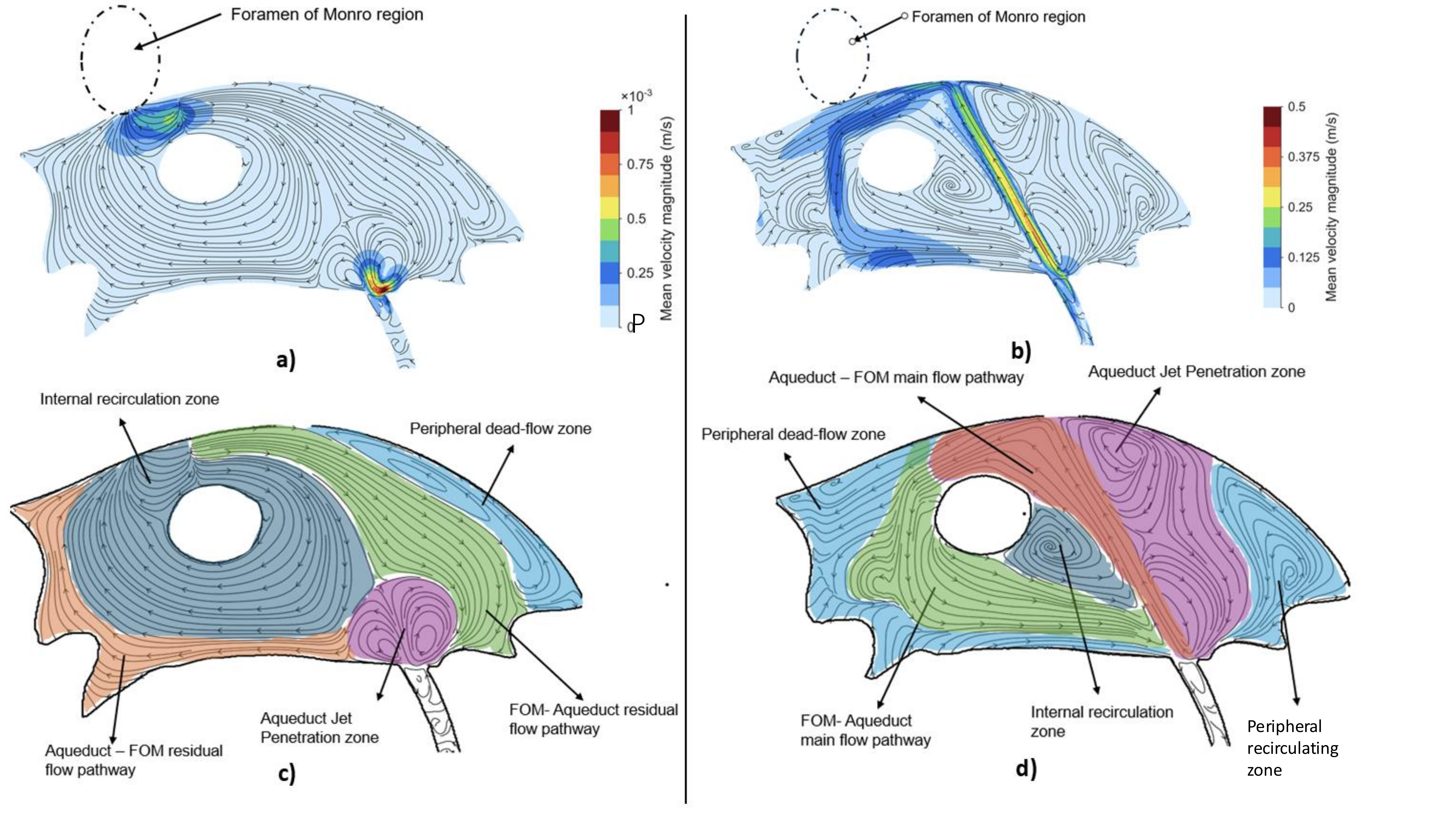}
\caption{Comparison of cycle-averaged midsagittal flow organisation of \(\widehat{\mathbf{U}}_s\) under (a,c) awake and (b,d) sleep-related conditions. Panels (a) and (b) compare the mean velocity magnitude and streamlines; separate colour-bar ranges are used because of the substantially different velocity scales. The dashed circles indicate the approximate projected regions of the FOM, which do not intersect the midsagittal plane directly. Panels (c) and (d) provide qualitative classifications of the corresponding mean streamline fields, identifying the FOM--aqueduct and aqueduct--FOM pathways, internal recirculation zone, aqueduct jet penetration zone, and peripheral dead-flow zones.}
\label{fig:midsagittal_mean_velocity}
\end{figure}

Having established the phase-dependent bidirectional motion, the cycle-averaged velocity fields were examined to determine whether the oscillatory flow generated an organised residual streaming pattern. The midsagittal mean fields showed clear differences between the awake and sleep conditions (Figure~\ref{fig:midsagittal_mean_velocity}). In the awake condition, the mean velocity magnitude remained low over most of the third ventricle, with higher values mainly localised near the FOM and cerebral aqueduct. The residual motion in the central ventricular cavity was therefore dominated by a large-scale circular motion around the adhesion at very low velocities.

Under sleep-related forcing, the mean velocity magnitude increased along the principal connection between the superior ventricular region and the aqueductal outlet. A more distinct jet-like structure appeared near the cerebral aqueduct, accompanied by stronger streamline curvature and local recirculation along the jet pathway. These cycle-averaged patterns show that the difference between the two conditions was not limited to the magnitude of the instantaneous oscillation: sleep-related forcing also generated a stronger and more clearly organised residual streaming field.

The annotated mean-streamline maps further classified the residual flow into FOM--aqueduct and aqueduct--FOM pathways, an internal recirculation zone, an aqueduct jet-penetration zone, and peripheral dead-flow zones (Figure~\ref{fig:midsagittal_mean_velocity}). These labels describe the organisation of the cycle-averaged field and should not be equated directly with the caudal and rostral directions of the instantaneous half-cycles. In the awake condition, both connections appeared as residual pathways around a broad internal recirculation zone which covered half of the ventricular space. In the sleep condition, they formed more clearly defined main pathways, with a distinct aqueduct jet-penetration zone and clearer separation from the peripheral dead-flow regions.

\subsection{Streaming across the aqueduct and quantitative flow metrics}

\begin{figure}[H]
\centering
\includegraphics[width=\textwidth]{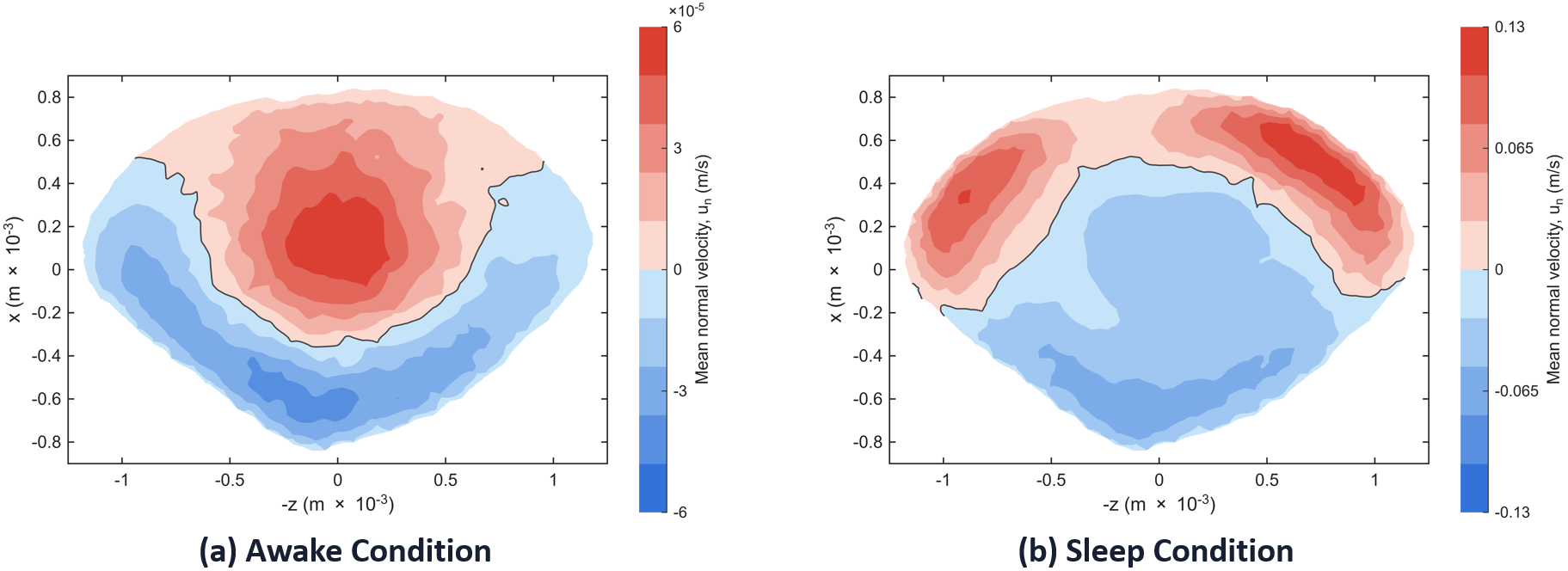}
\caption{Time-averaged normal velocity distributions of \(\widehat{\mathbf{U}}_s\) at the inferior end of the cerebral aqueduct under (a) awake and (b) sleep-related conditions. Red indicates positive caudal flow from the third ventricle towards the fourth ventricle, whereas blue indicates negative rostral flow towards the third ventricle. separate colour-bar ranges are used because of the substantially different velocity magnitudes.}
\label{fig:outlet_mean_velocity}
\end{figure}

To quantify the cross-ventricular transport associated with the streaming effect, the cycle-averaged velocity distribution was evaluated at the cerebral aqueduct outlet, see (Figure~\ref{fig:outlet_mean_velocity}). The cross-sectional shape has the bottom corner (z= 0,x=-0.008) facing anteriorly (ventrally) toward the front of the brain, representing the inner curvature wall of the aqueduct. In awake situation, this is where most of the flow goes rostral (negative velocity) into the 3rd ventricle, while at the center of the aqueduct flow is caudal. This changes in sleep when the center part flow is forced to reverse rostral, while the caudal flow regions are squeezed into the outer curvature walls of the aqueduct. The magnitude of the streaming velocity was substantially greater under sleep-related forcing. This opposing motion across the outlet cross-section is consistent with a residual streaming pattern superimposed on the predominantly reciprocating flow.

The dimensionless, exchange-related, and pressure metrics are summarised in Table~\ref{tab:flow_metrics}. 

\begin{table}[H]
\centering
\caption{Dimensionless metrics for the awake and sleep-related simulations. Here, \(Q_rT\) is the residual streaming volume accumulated over one cycle, \(SE=Q_rT/Q_s\) is the streaming efficiency, and \(\Delta p_{\mathrm{peak}}\) is the maximum absolute difference between the area-weighted mean pressures at the bilateral FOM and the cerebral aqueduct outlet.}
\label{tab:flow_metrics}
\resizebox{\textwidth}{!}{%
\begin{tabular}{lccccccc}
\toprule
Case &
\(Re_{\mathrm{Aq,peak}}\) &
\(\alpha\) &
\(SV\) \((\mu\mathrm{L/cycle})\) &
\(Q_rT\) \((\mu\mathrm{L/cycle})\) &
\(SE\) \((\%)\) &
\(\Delta p_{\mathrm{peak}}\) \((\mathrm{Pa})\) \\
\midrule
Awake & 12.7 & 2.38  & 0.0955 & \(5.12\times10^{-4}\) & 0.540 & 1.49 \\
Sleep & 1270  & 1.19 & 57.9  & 16.8                  & 28.4  & 792  \\
\bottomrule
\end{tabular}%
}
\end{table}

 The streaming efficiency \(SE=Q_rT/Q_s\) increased from \(0.540\%\) to \(28.4\%\). Thus, sleep-related forcing increased not only the magnitude of the reciprocating exchange but also the proportion expressed as residual streaming across the aqueduct.

\section{Discussion}

This study compared physiologically pressure-driven reciprocating CSF motion, represented by prescribed velocity boundaries, under awake and sleep-related forcing in the same representative human third-ventricle geometry. Sleep-related forcing produced greater velocity magnitude and outlet exchange, together with clearer caudal and rostral preferential pathways and a stronger aqueductal jet. While the nominal sleep stroke volume exceeded the reconstructed ventricular volume, this does not mean that there is replacement of the complete ventricular contents in each cycle. Rather,  streaming is to characterize cross-ventricular bidirectional exchange. The streaming efficiency measured in the aqueduct increased from \(0.54\%\) to \(28.4\%\), indicating that both the strength of oscillatory exchange and the organisation of the residual streaming flow pattern enhanced cross-ventricular transport.  

Figure~\ref{fig:midsagittal_mean_velocity} separates mean-flow organisation from instantaneous direction. The FOM--aqueduct and aqueduct--FOM labels identify connections within the mean streamline field, whereas caudal and rostral refer to opposite phases of the instantaneous oscillation. Flow moved from the FOM towards the aqueduct during the caudal phase and reversed during the rostral phase. In the awake condition, the connections remained residual pathways around a broad internal recirculation zone; under sleep-related forcing, they became more focused main pathways with clearer separation from the internal recirculation, jet-penetration, and peripheral dead-flow regions. Greater pathway continuity implies that a larger part of the cavity participated coherently in the cross-ventricular exchange. The forcing therefore reorganised the flow rather than merely scaling its velocity.

The stronger jet-like structure near the cerebral aqueduct provides a plausible fluid-mechanical explanation for this reorganisation. A higher-velocity aqueductal jet may entrain adjacent low-velocity CSF into the main pathway and thereby promote local recirculation and intraventricular mixing. This mechanism can transfer momentum between the narrow high-speed pathway and the weakly moving surrounding fluid without requiring a sustained net flow. The spatially non-uniform mean velocity at the outlet further indicates that this residual motion is organised across the section rather than behaving as a uniform plug flow. The sharper velocity gradients also suggest stronger shear-layer interaction. 

The streaming efficiency measured in the aqueduct increased from \(0.540\%\) to \(28.4\%\)  indicating that sleep-related forcing increased not only the reciprocating-flow magnitude but also the fraction represented by cross-ventricular streaming. This distinction is important because a stronger oscillation does not necessarily imply a proportionally larger residual component. 

It should be noted that the observed streaming efficiency is a hydrodynamic descriptor and cannot be interpreted directly as a measure of physiological clearance efficiency, which should be investigated in future work. However, our findings may suggest that streaming is relevant as a further mechanism that could support metabolite clearance \citep{xie2013sleep}, a mechanism not previously studied in this context. 
The present results add a third-ventricle-scale interpretation: stronger forcing may reorganise local pathways as well as increase cross-ventricular streaming exchange. 

The current study builds on previous CFD studies that have already examined third-ventricle CSF flow, aqueductal jets, recirculation, mixing, and mass transfer \citep{kurtcuoglu2007computational,kurtcuoglu2007mixing}. The present contribution is the controlled comparison of awake and sleep-related forcing within the same geometry, introducing pathway classification and streaming analysis. The results are specific for this selected geometry of the ventricle, including the position and size of the interthalamic adhesion. Additional assumption were made about the sleep-related forcing boundary conditions, derived from fMRI signals and reasonable arguments reported in \cite{fultz2019coupled}, while direct measurements of aqueductal velocities with sufficient resolution during NREM sleep were not available. 


\section{Conclusion}

This study compared cerebrospinal fluid (CSF) motion in one representative human third-ventricle geometry under oscillatory flow conditions representative of the pulse waves induced by the cardiac and respiratory cycles during wakefulness and NREM sleep. Sleep-related forcing imposed stronger bulk CSF displacement over the cycle; however this is not a simple scaling of the awake field. Rather, it showed the formation of organised caudal and rostral preferential pathways with a large increase in cross-ventricular exchange due to streaming. The streaming efficiency at the aqueduct increased from \(0.540\%\) to \(28.4\%\), indicating that streaming transport represented a considerable proportion of the cross-ventricular fluid exchange. Furthermore, the awake condition shows a larger recirculating region in the ventricle, which suggests weak mixing and fluid exchange. This changes substantially under sleep-induced forcing, leading to improved mixing and wash-out of dead-water zones.   


\section*{Funding}
The position of Christoph Bruecker is co-funded as the BAE SYSTEMS Sir Richard Olver Chair and the Royal Academy of
Engineering Chair (grant RCSRF1617/4/11). The work was partly supported in its initial stage by the Engineering and Physical Sciences Research Council (EPSRC) in DTP 2224. 

\section*{Acknowledgments}
This paper is dedicated to the memory of Luke McTavish. The authors thank Lady Dorit Young of Dartington for her interest and inspiration to further investigate flow in the human ventricles as part of the complex metabolic system of the brain.  

\section*{Declaration of competing interest}
The authors declare that they have no known competing financial interests or personal relationships that could have appeared to influence the work reported in this paper.

\section*{Data availability}
The data that support the findings of this study are available from the corresponding author upon reasonable request.

\section*{Declaration of generative AI and AI-assisted technologies in the manuscript preparation process}
During the preparation of this work, the authors used OpenAI ChatGPT/Codex to assist with language editing, formatting, and manuscript organisation. After using this tool, the authors reviewed and edited the content as needed and take full responsibility for the content of the publication.

\bibliographystyle{elsarticle-harv}
\bibliography{reference/references}

\appendix
\renewcommand{\thesection}{Appendix~\Alph{section}}
\renewcommand{\tablename}{Table}
\renewcommand{\figurename}{Figure}
\counterwithin*{table}{section}
\counterwithin*{figure}{section}
\counterwithin*{equation}{section}
\renewcommand{\thetable}{\Alph{section}.\arabic{table}}
\renewcommand{\thefigure}{\Alph{section}.\arabic{figure}}
\renewcommand{\theequation}{\Alph{section}.\arabic{equation}}

\section{}
\label{app:independence}

\subsection{Mesh independence}

A mesh-independence comparison was performed using the sleep case because it provided the more demanding test of spatial resolution. The baseline mesh contained approximately \(4.84\) million elements. A refined mesh increased the spatial resolution in all directions and contained approximately \(20\) million elements. The same post-processing workflow and 101 samples from the selected oscillation cycle were used for both datasets. The comparison focused on stroke volume, residual streaming flux, stream-effect ratio, peak outlet flow, and representative maximum velocities.

\begin{table}[htbp]
\centering
\caption{Mesh-independence comparison for the sleep simulation. The relative difference was calculated with respect to the baseline-mesh result; \(Q_r\) denotes residual streaming flux.}
\label{tab:mesh_independence}
\small
\resizebox{\textwidth}{!}{%
\begin{tabular}{@{}lrrrr@{}}
\toprule
Metric & Baseline mesh & Refined mesh & Difference & Difference (\%) \\
\midrule
\(SV\) \((\mu\mathrm{L/cycle})\) & 2432.117 & 2432.667 & 0.549 & 0.023 \\
\(Q_rT\) \((\mu\mathrm{L/cycle})\) & 547.634 & 532.003 & -15.630 & -2.854 \\
\(SE\) \((\%)\) & 22.513 & 21.869 & -0.644 & -2.860 \\
\bottomrule
\end{tabular}%
}
\end{table}

The integrated outlet-flow quantities showed very small differences between the two datasets. Stroke volume differed by \(0.023\%\), while the peak positive and negative outlet flow rates differed by \(0.004\%\) and \(0.038\%\), respectively. The principal cycle-integrated exchange quantities were therefore insensitive to mesh refinement.

The residual streaming flux \(Q_r\) showed a larger difference of \(-2.854\%\), and \(Q_r/Q_s\) changed from \(22.513\%\) to \(21.869\%\), an absolute change of \(0.644\) percentage points. This greater sensitivity is expected because \(Q_r\) is derived from the positive part of the residual mean field and therefore depends on local velocity structure. The midsagittal maximum mean velocity differed by \(1.054\%\), supporting adequate resolution of the principal intraventricular velocity scale.

Overall, the mesh-independence comparison showed that the selected mesh provided stable predictions of the principal outlet exchange metrics, stroke volume, stream-effect ratio, and representative velocity scales. The mesh was therefore considered sufficiently resolved for the comparative awake and sleep-related simulations.
\subsection{Time-step independence}

The time-step selection was assessed using the representative CFL estimates reported in the numerical method. The awake and sleep simulations gave \(C_{\mathrm{awake}}\approx0.38\) and \(C_{\mathrm{sleep}}\approx0.94\), respectively, indicating that the selected solver time steps were adequate for the prescribed peak velocities and representative mesh length.

\subsection{Cycle independence}

A cycle-independence test was performed for the sleep simulation by comparing the second and third oscillation cycles. The sleep case was used because it produced the strongest velocity magnitude and largest exchange volume, and therefore provided the most stringent test of temporal periodicity. The same post-processing procedure was applied to both cycles, and the comparison included outlet exchange volume, stroke volume, net flow, and peak outlet flow rates.

\begin{table}[htbp]
\centering
\caption{Cycle-independence comparison between the second and third sleep cycles. The percentage difference was calculated with respect to Cycle 2.}
\label{tab:cycle_independence}
\small
\resizebox{\textwidth}{!}{%
\begin{tabular}{@{}lrrrr@{}}
\toprule
Metric & Cycle 2 & Cycle 3 & Difference & Difference (\%) \\
\midrule
Inflow magnitude \((\mu\mathrm{L/cycle})\)
& 2431.706 & 2431.719 & 0.01249 & 0.000514 \\

\(SV\) \((\mu\mathrm{L/cycle})\)
& 2432.112 & 2432.117 & 0.00496 & 0.000204 \\

\(Q_rT\) \((\mu\mathrm{L/cycle})\)
& 540.670 & 549.803 & 9.13295 & 1.68919 \\

\bottomrule
\end{tabular}%
}
\end{table}

The comparison showed that the main cycle-integrated flow quantities were essentially unchanged between the second and third cycles. Stroke volume differed by \(0.000204\%\), while the difference in streaming transport was only \(1.7\%\), which is low compared with the total streaming flux exchange. 
The third cycle was therefore considered to represent a cycle-independent periodic state and was used for the reported sleep-flow analysis.
\section{Estimation of the sleep inlet velocity}
\label{app:sleep_velocity_estimation}

The sleep-related forcing data were approximated from the fMRI data presented by \citet{fultz2019coupled}. Their NREM sleep sequence, shown in Figure 2D, illustrates successive waves of large CSF inflow-related signal changes (pulse trains of 3--4 pulses), from which we derived an approximate pulse-time interval of \(T \approx 4\,\mathrm{s}\). Further information was obtained from their video in the supplementary material, published in the PubMed Central (PMC) repository. The blue region shown in the fMRI colour map covers a large area in the fourth ventricle during the sudden evacuation of blood; therefore, the resulting drop in pressure is assumed to act uniformly across the fourth ventricle, drawing the CSF upwards as a coordinated liquid plug. As reported in their paper, during some of those events, the velocity considerably exceeded the critical imaging velocity of \(11.4\,\mathrm{mm\,s^{-1}}\) for slice 2, which is located in the fourth ventricle and corresponds to the deep-blue region in their movie during one of the strong flushing events. For the simplified CFD condition, a representative maximum velocity in the fourth ventricle was approximated as \(20\,\mathrm{mm\,s^{-1}}\). Scaling this enhanced inflow to the bilateral FOM in the reconstructed model gave a prescribed inlet amplitude of \(U_{\mathrm{peak},\mathrm{sleep}}=0.17\,\mathrm{m\,s^{-1}}\). The resulting nominal positive-half-cycle stroke volume was \(2.43\,\mathrm{mL}\).

\end{document}